\documentclass[prl,showpacs,twocolumn,aps]{revtex4}
\usepackage{graphicx}
\newcommand{\be}{\begin{equation}}
\newcommand{\ee}{\end{equation}}
\newcommand{\bea}{\begin{eqnarray}}
\newcommand{\eea}{\end{eqnarray}}

\newcommand{\bwt}{\begin{widetext}}
\newcommand{\ewt}{\end{widetext}}
\begin{document}
\title{Microwave Conductivity of a High Purity $d$-wave Superconductor
}
\author{Wonkee Kim$^{1}$, F. Marsiglio$^{1}$, and J.P. Carbotte$^{2}$}
\affiliation{
$^{1}$Department of Physics, University of Alberta, Edmonton, Alberta,
Canada, T6G~2J1
\\
$^{2}$ Department of Physics and Astronomy, McMaster University,
Hamilton, Ontario, Canada}
\begin{abstract}
The cusp-like behavior of the microwave conductivity observed in
clean ortho-II YB$_{2}$Cu$_{3}$O$_{6.50}$ at 
low temperature and low frequency
is shown to be related directly to a linear in frequency dependence of the
impurity scattering rate with a negligibly small value at zero frequency.
In the weak scattering limit, the conductivity decreases linearly
with the frequency. In the vortex state, assuming a random (Gaussian)
distribution of vortices, we show that the magnetic field profoundly
alters the impurity scattering rate, which now acquires a finite zero 
frequency value. As a consequence, we predict
a Drude-like line shape in the 
microwave conductivity at low frequency.  
\end{abstract}

\date{\today}
\maketitle

At low temperature elastic impurity scattering is expected to
dominate over its inelastic counterpart in high-$T_{c}$ 
cuprates\cite{bonn,hosseini,nuse}.
In much of the early literature on impurity scattering in $d$-wave 
superconductors, only the unitary (strong) and Born (weak) limit
were studied\cite{hirschfeld1,hirschfeld2}.
Later, it was recognized that neither of these two limits
apply and an intermediate impurity potential $V_{i}$
is realistic for the high purity samples which are now available. 
In particular, Schachinger and Carbotte\cite{ewald1} found that
a finite value of $c=1/[\pi N(0)V_{i}]$ with $N(0)$ the density of states
in the normal state, is required to qualitatively fit the microwave data
of ortho-I YB$_{2}$Cu$_{3}$O$_{6.99}$. Similarly,
intermediate scattering was necessary to explain the thermal conductivity
of another high purity ortho-I sample\cite{hill,kim}. 
Additional evidence for a finite value of the scattering strength $c$
is the zero temperature residual absorption\cite{ewald2} 
observed in ortho-II 
YB$_{2}$Cu$_{3}$O$_{6.50}$ while
at the same time
the penetration depth is linear in $T$ and 
the microwave conductivity at low frequency 
exhibits a cusp\cite{turner,harris}.

We consider the microwave conductivity of an ortho-II
sample with and without a magnetic field $(H)$.
Our theoretical results with $c=0.4$ semi-quantitatively explain the 
experimental data in zero field. We do not attempt an exact
fit to the data, instead we explore the origin of the cusp behavior
of the microwave conductivity at low frequency and changes resulting
from the application of a magnetic field.
We find that the cusp arises because
the impurity scattering rate 
is the form
$\gamma(\omega)\simeq\gamma_{00}+\alpha\omega$
for small $\omega$ with $\gamma_{00}$ negligibly small. It is the linear
increase of $\gamma(\omega)$ that is responsible for the cusp behavior.
In the vortex state, the impurity scattering is profoundly changed,
and is no longer linear.
Moreover, its zero frequency value is increased significantly.
Consequently, we predict that
the cusp disappears and a Drude-like behavior results.

Roughly speaking, the magnetic field has two independent effects on the
conductivity. It increases the number of quasiparticles, and so
the conductivity. The other is its effect on the
impurity scattering. This is the more interesting and significant 
aspect for understanding
quasiparticle transport in $d$-wave superconductors.
The formalism to account for field effects
on physical quantities such as thermodynamic and transport properties
is well established.
The basic approach uses a semiclassical approximation\cite{volovik,kubert} 
in which
the essential effect of the field is embodied via the Doppler shift
due to the circulating supercurrent $(v_{s})$ around the vortex cores. 
This approximation is valid when $H_{c1}\lesssim H\ll H_{c2}$.
Another approximation valid for the low $T$ microwave conductivity
is that the nodal quasiparticles dominantly
determine the transport properties.

Due to the inhomogeneous nature of the vortex state, the impurity scattering
$\gamma(\omega,{\bf r})$
depends on local position $({\bf r})$ in space. Since the microwave
conductivity is significantly affected by the impurity scattering,
it is quite crucial to determine the scattering rate
self-consistently. The self-consistency comes through the T-matrix
approximation in which the local $2\times2$ Green's function 
${\hat G}({\bf k},i{\tilde\omega}-v_{s}({\bf r})\cdot{\bf k})$
appears with 
${\tilde\omega}=\omega-\Sigma_{i}({\tilde\omega},{\bf r})$
and $\gamma(\omega,{\bf r})=-\mbox{Im}\Sigma_{i,ret}({\tilde\omega},{\bf r})$.
The local impurity self-energy is determined by
\be
\Sigma_{i,ret}({\tilde\omega},{\bf r})=\frac
{\Gamma G_{0}({\tilde\omega},{\bf r})}{c^{2}-G^{2}_{0}({\tilde\omega},{\bf r})}\;,
\label{Sigma_r}
\ee
where $G_{0}({\tilde\omega},{\bf r})=[2\pi N(0)]^{-1}
\sum_{\bf k}\mbox{Tr}[{\hat G}_{ret}
({\bf k},i{\tilde\omega}-v_{s}({\bf r})\cdot{\bf k})]$ and $\Gamma$ is proportional 
to the impurity concentration.
Since the nodal quasiparticles dominantly contribute to transport at low T, 
the Doppler shift effectively depends only on momentum along the nodes; 
namely, 
$\epsilon_{i}={\bf v}_{s}({\bf r})\cdot{\bf k}_{ni}$, where $i$ is a node index.
Then,
\be
G_{0}(\tilde{\omega},\epsilon_{1},\epsilon_{2})\simeq
\frac{1}{2\pi}\sum_{i=1}^{4}
\frac{\zeta_{i}}{\Delta_{0}}\ln\left(\frac{\zeta_{i}}{4i\Delta_{0}}\right)\;
\ee
where 
$\zeta_{i}=\tilde{\omega}-\epsilon_{i}$ with
$\epsilon_{3}=-\epsilon_{1}$ and $\epsilon_{4}=-\epsilon_{2}$.
Now $\tilde{\omega}=\omega-\Sigma_{i}
(\tilde{\omega},\epsilon_{1},\epsilon_{2})$ and
$\gamma({\tilde\omega},\epsilon_{1},\epsilon_{2})=
-\mbox{Im}\Sigma_{i}(\tilde{\omega},\epsilon_{1},\epsilon_{2})$.
Consequently, only two nodes (say, node 1 and node 2) need to be 
considered independently in the self-consistent calculation\cite{vekhter} for 
$\Sigma_{i,ret}({\tilde\omega},\epsilon_{1},\epsilon_{2})$. Once the self-energy
is determined, the local microwave conductivity 
$\sigma_{xx}(\Omega, T, \epsilon_{1},\epsilon_{2})$ can be evaluated. However, it is
still necessary to average the conductivity over space to compare 
with experimental data:
\be
\sigma_{xx}(\Omega,T,H)=
\int{}d\epsilon_{1}d\epsilon_{2}\;{\cal L}(\epsilon_{1},\epsilon_{2})
\;\sigma_{xx}(\Omega,T,\epsilon_{1},\epsilon_{2})\;,
\ee
where ${\cal L}(\epsilon_{1},\epsilon_{2})$ 
is the vortex distribution function taking into account the two nodes.
As in the standard formalism, the optical conductivity is given by the spectral function 
corresponding to the $2\times2$ Green's function of the vortex state. 
Later, we will present the explicit form of the microwave conductivity. 
However, from the theoretical point of view, it is practically 
a formidable task. Here we reduce it to a
simpler, if approximate, form valid for low 
$T$ and for a wide range of the magnetic field.

We begin with a discussion of the zero field case. 
Microwave data has been obtained at various $T$ by Turner 
{\it et al.} \cite{turner}
for high purity samples of ortho-I YB$_{2}$Cu$_{3}$O$_{6.99}$ 
and ortho-II YB$_{2}$Cu$_{3}$O$_{6.50}$. 
Here we consider only the ortho-II sample. 
For $H=0$, the microwave conductivity 
$(\sigma_{xx}\equiv\sigma)$
is given by
\bwt
\be
\sigma(\Omega,T)=\frac{e^{2}}{2\pi^{2}}\left(\frac{v_{f}}{v_{g}}\right)
\int{}d\omega\left[\frac{f(\omega)-f(\omega+\Omega)}{\Omega}\right]
{\cal A}_{\sigma}(\Omega,\omega)\;,
\label{micro_zero_H}
\ee
\ewt
where
$
{\cal A}_{\sigma}(\Omega,\omega)=2(a+b+c)/d
$
with
\bea
a&=&\gamma\gamma_{\Omega}
\left[\bar{\omega}^{2}_{\Omega}-\bar{\omega}^{2}+
\gamma^{2}_{\Omega}-\gamma^{2}\right]
\ln\left[\frac{\bar{\omega}^{2}_{\Omega}+\gamma^{2}_{\Omega}}
{\bar{\omega}^{2}+\gamma^{2}}\right]
\nonumber\\
b&=&2\gamma\left[
\bar{\omega}_{\Omega}\left(\gamma^{2}_{\Omega}+\gamma^{2}\right)
-2\bar{\omega}\gamma^{2}_{\Omega}+\bar{\omega}_{\Omega}
\left(\bar{\omega}_{\Omega}-\bar{\omega}\right)^{2}\right]
\tan^{-1}\left(\frac{\bar{\omega}_{\Omega}}{\gamma_{\Omega}}\right)
\nonumber\\
c&=&2\gamma_{\Omega}\left[
\bar{\omega}
\left(\gamma^{2}_{\Omega}+\gamma^{2}\right)-
2\bar{\omega}_{\Omega}\gamma^{2}
+\bar{\omega}\left(\bar{\omega}_{\Omega}-\bar{\omega}\right)^{2}\right]
\tan^{-1}\left(\frac{\bar{\omega}}{\gamma}\right)
\nonumber\\
d&=&\left[
\left(\bar{\omega}_{\Omega}-\bar{\omega}\right)^{2}+
\left(\gamma_{\Omega}-\gamma\right)^{2}\right]
\left[
\left(\bar{\omega}_{\Omega}-\bar{\omega}\right)^{2}+
\left(\gamma_{\Omega}+\gamma\right)^{2}\right]\;,
\nonumber
\eea
here $\bar{\omega}=\omega-\mbox{Re}\left[\Sigma_{i,ret}(\omega)\right]$
while $\bar{\omega}_{\Omega}=\omega+\Omega-\mbox{Re}
\left[\Sigma_{i,ret}(\omega+\Omega)\right]$.
The impurity scattering rate
$\gamma(\omega)$ is given by the imaginary part of the self-energy, 
$-\mbox{Im}\Sigma_{i,ret}(\omega)$ and $\gamma_{\Omega}$
denotes $\gamma(\omega+\Omega)$. The self-consistent equation determining 
$\Sigma_{i,ret}$ is now 
$\Sigma_{i}({\tilde\omega})=\frac
{\Gamma G_{0}({\tilde\omega})}{c^{2}-G^{2}_{0}({\tilde\omega})}$,
where
$G_{0}(\tilde{\omega})\simeq
\frac{2}{\pi}
\frac{\tilde{\omega}}{\Delta_{0}}
\ln\frac{\tilde{\omega}}{4i\Delta_{0}}$
with $\tilde{\omega}=\omega-\Sigma_{i}({\tilde\omega})$ and
$\gamma(\tilde{\omega})=-\mbox{Im}\Sigma_{i}(\tilde{\omega})$.

An extensive search of parameter space for $c$ and $\Gamma$ has 
yielded a reasonable fit to the zero field microwave data of 
the ortho-II sample at $T=1.3$, $2.7$,
$4.3$, and $6.7$K, which is shown in Fig.~1. The parameters are
$c=0.4$ and $\Gamma/\Delta_{0}=0.003$.   
The solid curves are the results of our calculations while symbols are 
for the experimental data of Ref.\cite{turner}. 
For the comparison we assumed 
$\sigma_{00}\simeq2.5\times
10^{5}$ Ohm$^{-1}$m$^{-1}$. The self-consistently obtained $\gamma(\omega)$
is shown in the inset of Fig. 1. 
The curve peaks around $\omega/\Delta_{0}=0.2$ and 
has maximum height $\gamma(\omega)/\Delta_{0}\simeq0.008$. However, at low temperature 
only the very small $\omega$ region of $\gamma(\omega)$
plays a significant role in the calculation of $\sigma(\Omega, T)$. In this region,
we can approximate $\gamma(\omega)\simeq\gamma_{00}+\alpha\omega$ 
with a value of $\gamma_{00}$
which is practically zero, and $\alpha\simeq0.021$. 
As we will see, what is crucial for a cusp to appear in the microwave conductivity 
at low frequency is that $\gamma_{00}\ll T$ at the temperature considered
and a linear increase of $\gamma(\omega)$ at
small $\omega$. Moreover, at low frequency the conductivity is 
determined mainly by $\alpha$ and not by $\gamma_{00}$. 
This implies that the cusp-like behavior in highly pure samples is robust.

In the weak impurity scattering regime $(\gamma(w)<T)$, 
we can simplify the equation for the microwave conductivity significantly 
and obtain;
\be
\frac{\sigma(\Omega,T)}{\sigma_{00}}\simeq
\int{}d\omega\left(-\frac{\partial f}{\partial\omega}\right)
\pi|\omega|\frac{2\gamma(\omega)}{\Omega^{2}+4\gamma^{2}(\omega)}\;.
\label{simple_micro_zero_H}
\ee
If we use the above mentioned analytic form of $\gamma(\omega)$ 
in Eq.~(\ref{simple_micro_zero_H}), we obtain
remarkably good agreement with the full numerical calculation 
based on Eq.(\ref{micro_zero_H}). 
It is clear that the complicated behavior
of the microwave conductivity observed simply reflects 
the $\omega$ dependence of the underlying 
impurity scattering rate $\gamma(\omega)$. 
Consequently, instead of making use of the self-consistent
calculation for $\gamma(\omega)$, one could simply use a fitting procedure 
based on Eq.~(\ref{simple_micro_zero_H}) to get
the impurity scattering rate which gives the best fit to the experimental data.

When $\gamma_{00}<\Omega\ll T$, we can further simplify 
Eq.~(\ref{simple_micro_zero_H}).
The analytic expression we obtained, which is valid at small $\Omega(<T)$, is
\be
\frac{\sigma}{\sigma_{00}}\simeq
\frac{\pi}{2\alpha}\left[1-\frac{\pi}{8\alpha}\frac{\Omega}{T}
\right]\;.
\label{micro_analytic_zero_H}
\ee
This equation is one of our important results and explains the 
observed cusp-like behavior. 
As the dashed line in Fig.~1 shows for $T=1.3K$,
the dc conductivity is determined only 
by the slope $\alpha$, and also we see from
Eq.~(\ref{micro_analytic_zero_H}), it is independent of $\gamma_{00}$. 
Further, the value of the dc conductivity
is independent 
of $T$. The full numerical calculation shows a slight $T$ 
dependence particularly at $T=6.7$K.
This can be traced to the fact that the approximate expression
for $\gamma(\omega)$ is no longer completely valid.
Eq.(\ref{micro_analytic_zero_H}) shows that 
at low frequency, the microwave conductivity decreases linearly with $\Omega$, 
and the slope of the decrease is inversely proportional 
to $T$. This agrees well with experimental observation.
Consequently, the cusp arises because $\gamma_{00}$ is negligibly small
and $\gamma(\omega)$ increases linearly at small $\omega$. 
No Drude fit is possible for the highly pure ortho-II sample in zero field. 
However, we will show that the $\omega$ dependence of the impurity scattering 
rate is profoundly changed in the presence of a magnetic field. 
As a consequence, the shape of the microwave 
conductivity as a function of
frequency is also fundamentally altered.     
\begin{figure}
\begin{center}
\includegraphics[height=2.6in,width=3.1in]{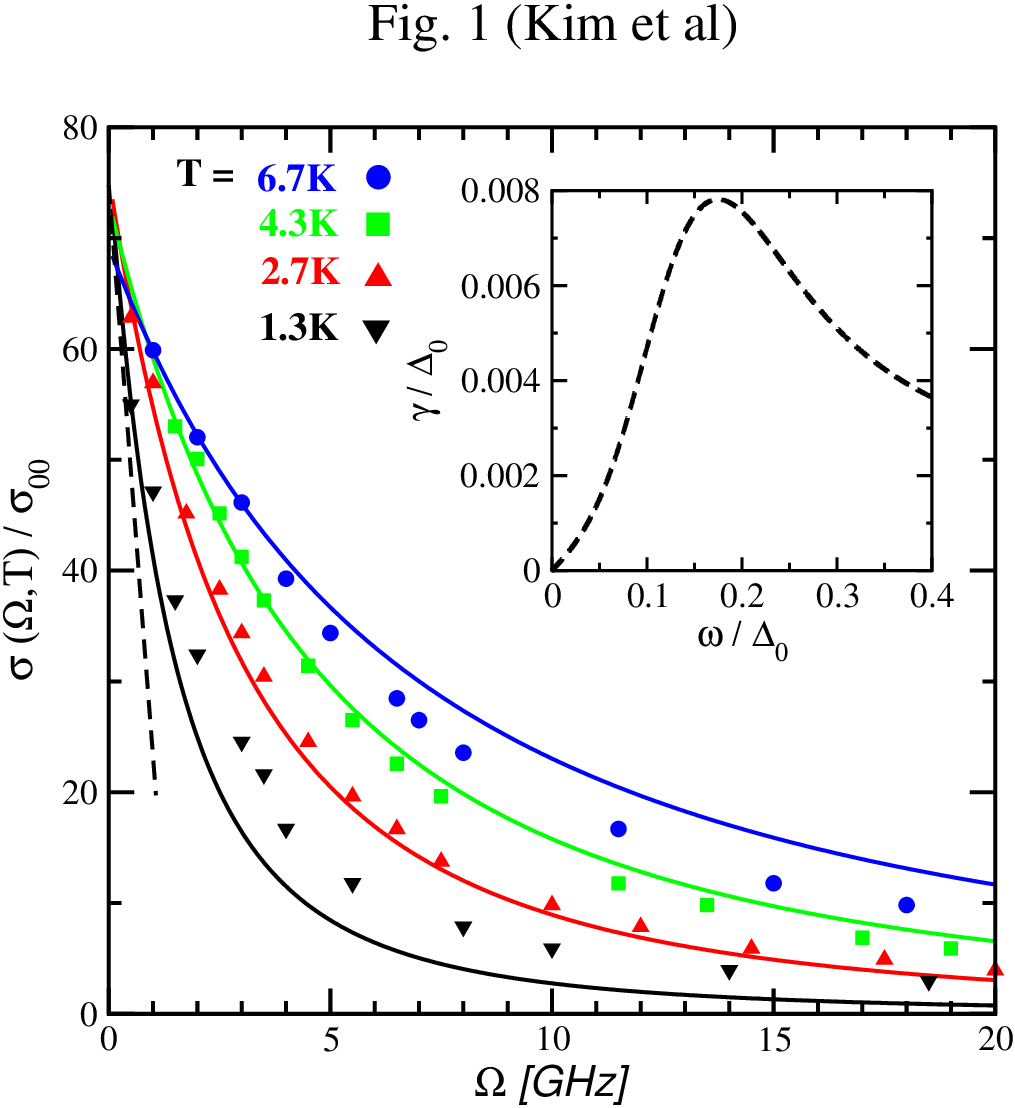}
\caption{Microwave conductivity $\sigma(T,\Omega,H=0)$ vs. $\Omega$
up to $20$GHz. Data of Ref.\cite{turner} are shown for
$T=1.3$(triangle down), $2.7$(triangle up), $4.3$(square), and
$6.7$K(circle). The solid curves are our numerical results using
Eq.~(\ref{micro_zero_H}).
The dashed line based on the analytic expression
Eq.~(\ref{micro_analytic_zero_H}) (lowest temperature)
shows the cusp behavior at low frequency.
The parameters for impurity scattering
are $\Gamma/\Delta_{0}=0.003$ and $c=0.4$. The inset gives
the self-consistently obtained
impurity scattering rate $\gamma(\omega)$.
}
\end{center}
\end{figure}

Next we consider the microwave conductivity in the vortex state.
As explained earlier, we need to determine $\gamma(\omega,\epsilon_{1},\epsilon_{2})$
to evaluate $\sigma(\Omega,T,\epsilon_{1},\epsilon_{2})$. Then an averaging procedure
is necessary to obtain $\sigma(\Omega,T,H)$, which can be compared with experimental
data. To get the formal expression for $\sigma(\Omega,T,\epsilon_{1},\epsilon_{2})$, 
we only need to replace ${\cal A}_{\sigma}(\Omega,\omega)$ in
Eq.~(\ref{micro_zero_H}) with 
$\sum_{i=1}^{2}{\cal A}_{\sigma,i}(\Omega,\omega,\epsilon_{1},\epsilon_{2})$, where
$i$ is again the node index. For the node $i$, ${\bar\omega}\rightarrow
\omega-\mbox{Re}\left[\Sigma_{i,ret}
(\omega,\epsilon_{1},\epsilon_{2})\right]-\epsilon_{i}$
and the corresponding 
$\gamma(\omega)\rightarrow\gamma(\omega,\epsilon_{1},\epsilon_{2})$.
Obviously, such a procedure requires a huge numerical calculation.
Because of this it is important to find an
alternative, if approximate, way to evaluate 
$\sigma(\Omega,T,H)$. We base our simplification on the physical fact
that the field has two effects on the conductivity.
The first is to create quasiparticles.
The second is to alter the impurity scattering itself, and this
will significantly change the shape of the microwave conductivity as a function of
frequency. In our alternative way, we treat the two effects separately. 
First we deal with
the field effect on the impurity self-energy to obtain $\Sigma_{i,ret}(\omega,H)$. 
Then,
we take into account the increase in the number of quasiparticles. 
This procedure simplifies
calculations of the microwave conductivity considerably. It reduces the contributions
from the quasiparticles in the two distinct nodal areas to one from 
a single average node. 
From the theoretical point of view, this means that 
the space-averaged Green's function
${\hat G}({\bf k},i{\tilde\omega},H)$ is used in the T-matrix approximation to obtain
\be
\Sigma_{i}({\tilde\omega},H)=\frac
{\Gamma G_{0}({\tilde\omega},H)}{c^{2}-G^{2}_{0}({\tilde\omega},H)}\;,
\label{self_imp_H}
\ee
where
$
G_{0}(\tilde{\omega},H)\simeq
\frac{2}{\pi}\int{}d\epsilon\;
{\cal P}(\epsilon)
\frac{z}{\Delta_{0}}\ln\frac{z}{4i\Delta_{0}}
$
where $z=\tilde{\omega}-\epsilon$
and ${\cal P}(\epsilon)=\int{}d\epsilon'\;{\cal L}(\epsilon,\epsilon')$.

Within this approximation, the only change needed 
in Eq.(\ref{micro_zero_H}) is to replace ${\cal A}_{\sigma}(\Omega,\omega)$ with
${\cal A}_{\sigma}(\Omega,\omega,\epsilon,H)$ and the averaging procedure requires
the single node distribution of vortices ${\cal P}(\epsilon)$ as already indicated
in the self-consistent calculation for the impurity self-energy.
We have tested the single-node approximation
involving ${\cal A}_{\sigma}(\Omega,\omega,\epsilon,H)$ with
the two-node method using 
${\cal A}_{\sigma,i}(\Omega,\omega,\epsilon_{1},\epsilon_{2})$,
on the dc conductivity at zero temperature, for which
only $\gamma(0,\epsilon_{1},\epsilon_{2})$ or $\gamma(0,H)$ enters
the calculation. We found that these two methods
give almost identical results for $0<E_{H}/\Delta_{0}\le0.1$ when we consider a
random (Gaussian) distribution of vortices \cite{other_dist}. 
Therefore, as a first approximation, we use the simple procedure of 
Eq.~(\ref{self_imp_H}) to evaluate
the low temperature microwave conductivity.
\begin{figure}
\begin{center}
\includegraphics[height=2.6in,width=3.1in]{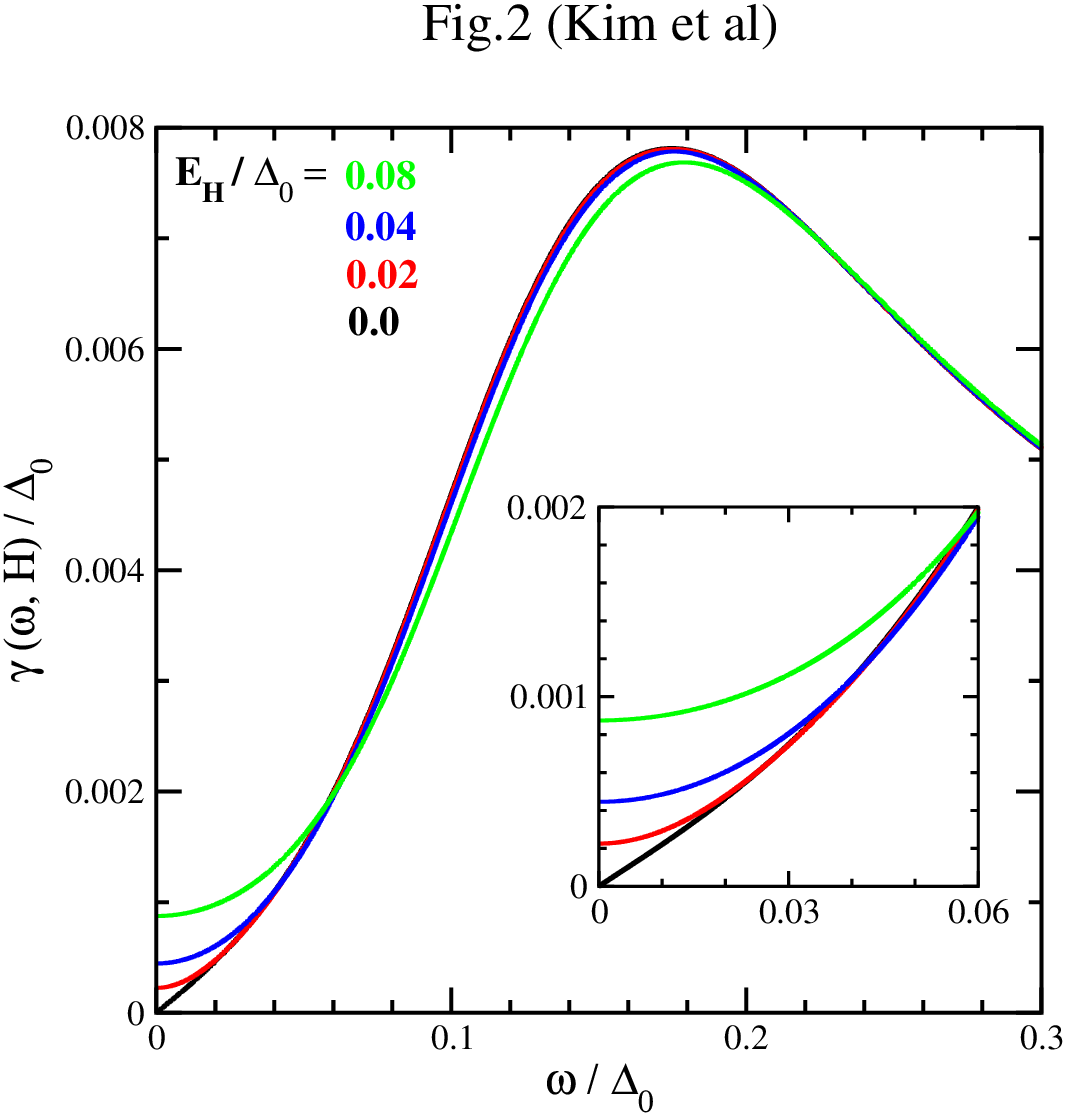}
\caption{Impurity scattering rate $\gamma(\omega,H)$
for $E_{H}/\Delta_{0}=0$, $0.02$, $0.04$, and $0.08$.
Strong changes are seen at small $\omega$.
The inset is for the small $\omega$ range.
It is clear from the inset that $\gamma(\omega,H)$
changes quadratically and $\gamma(0,H)$ is
no longer negligible.
}
\end{center}
\end{figure}

In Fig.~2, we show our
results for $\gamma(\omega,H)$ 
as a function of $\omega$
obtained self-consistently
for various values of the magnetic energy $E_{H}/\Delta_{0}=0.02$, 
$0.04$, and $0.08$.
We estimate 
$E_{H}\simeq 20\sqrt{H}KT^{-1/2}$ with the 
magnetic field in units of Tesla.
The same values of the parameters $c$ and $\Gamma$ are used as before.
For $E_{H}/\Delta_{0}=0$, the result for $\gamma(\omega,H)$ is the same as shown in the 
inset of Fig.~1. As one can
see, the field has a drastic effect on the low frequency 
region of the impurity scattering, 
and this is the region that is
mainly sampled in the low $T$ microwave conductivity. The two most important changes in 
$\gamma(\omega,H)$ are i) $\gamma(0,H)$ is no longer negligible and ii) for low $\omega$,
the increase in $\gamma(\omega,H)$ is now quadratic:
$\gamma(\omega,H)/\Delta_{0}\simeq\gamma(0,H)/\Delta_{0}
+\beta\left(\omega/\Delta_{0}\right)^{2}$, where $\beta\simeq0.6$, $0.4$, and $0.26$
for $E_{H}/\Delta_{0}=0.02$, $0.04$, and $0.08$, respectively.
The results for the microwave conductivity $\sigma(\Omega,T,H)$ are shown in Fig.~3 
up to $20GHz$ at $T=1.3K$. The black solid curve $(E_{H}/\Delta_{0}=0)$ is 
repeated from Fig.~1 for comparison. The other curves correspond to 
$E_{H}/\Delta_{0}=0.02$ (red), $0.04$ (blue), and $0.08$ (green), respectively. 
The magnetic field removes the cusp at low frequency. 
This is due to the drastic change 
in impurity scattering in the vortex state as we will show below.

Even in the vortex state, we can still simplify the equation for the microwave
conductivity in the weak scattering limit if $\gamma(w,H)<T$ for the important
range of $\omega$. We obtain
\be
\frac{\sigma}{\sigma_{00}}\simeq
\int{}d\omega\left(-\frac{\partial f}{\partial\omega}\right)
\pi {\cal N}(\omega,H)\frac{2\gamma(\omega,H)}
{\Omega^{2}+4\gamma^{2}(\omega,H)}\;,
\label{micro_vortex}
\ee
where the effective density of states is
${\cal N}(\omega,H)=\int d\epsilon\;{\cal P}(\epsilon)|\omega-\epsilon|$.
Eq.~(\ref{micro_vortex}) clearly shows both the
effect of the magnetic field on the density of states and on the impurity
scattering. The symbols in Fig.~3 are based on Eq.~(\ref{micro_vortex}) 
with the approximate quadratic expression for 
$\gamma(w,H)$ quoted. It is clear that the microwave conductivity 
is well represented by the weak impurity scattering limit 
even in the vortex state.
Moreover, from the comparison with Eq.(\ref{simple_micro_zero_H}), 
we can see immediately
the field effect in Eq.(\ref{micro_vortex}). It represents the appropriate
generalization
of Eq.(\ref{simple_micro_zero_H}), and is another of our important results.
For the random vortex distribution, the effective density of states becomes
${\cal N}(\omega,H)=\left({E_{H}}/{\sqrt{\pi}}\right)e^{-\omega^{2}/E^{2}_{H}}+
\omega\;\mbox{erf}\left({\omega}/{E_{H}}\right)$.
In the field dominated regime, we can further simplify Eq.~(\ref{micro_vortex}) 
utilizing the fact that $\gamma(\omega,H)\approx\gamma(0,H)$ and 
${\cal N}(\omega,H)\approx E_{H}/\sqrt{\pi}$ at very low 
$\omega\ll E_{H}$. Now it is possible to approximate the microwave conductivity as:
\be
\frac{\sigma(\Omega,T,H)}{\sigma_{00}}\approx\frac{\sqrt{\pi}E_{H}}
{2\gamma(0,H)}\left[1-\frac{\Omega^{2}}{4\gamma^{2}(0,H)}\right]\;.
\label{drude_like}
\ee
This equation shows that the microwave conductivity 
now has no cusp and a Drude-like behavior appears in the vortex state 
which can be traced to the characteristic change in the impurity scattering
which now has a finite $\gamma(0,H)$ value.
\begin{figure}
\begin{center}
\includegraphics[height=2.6in,width=3.1in]{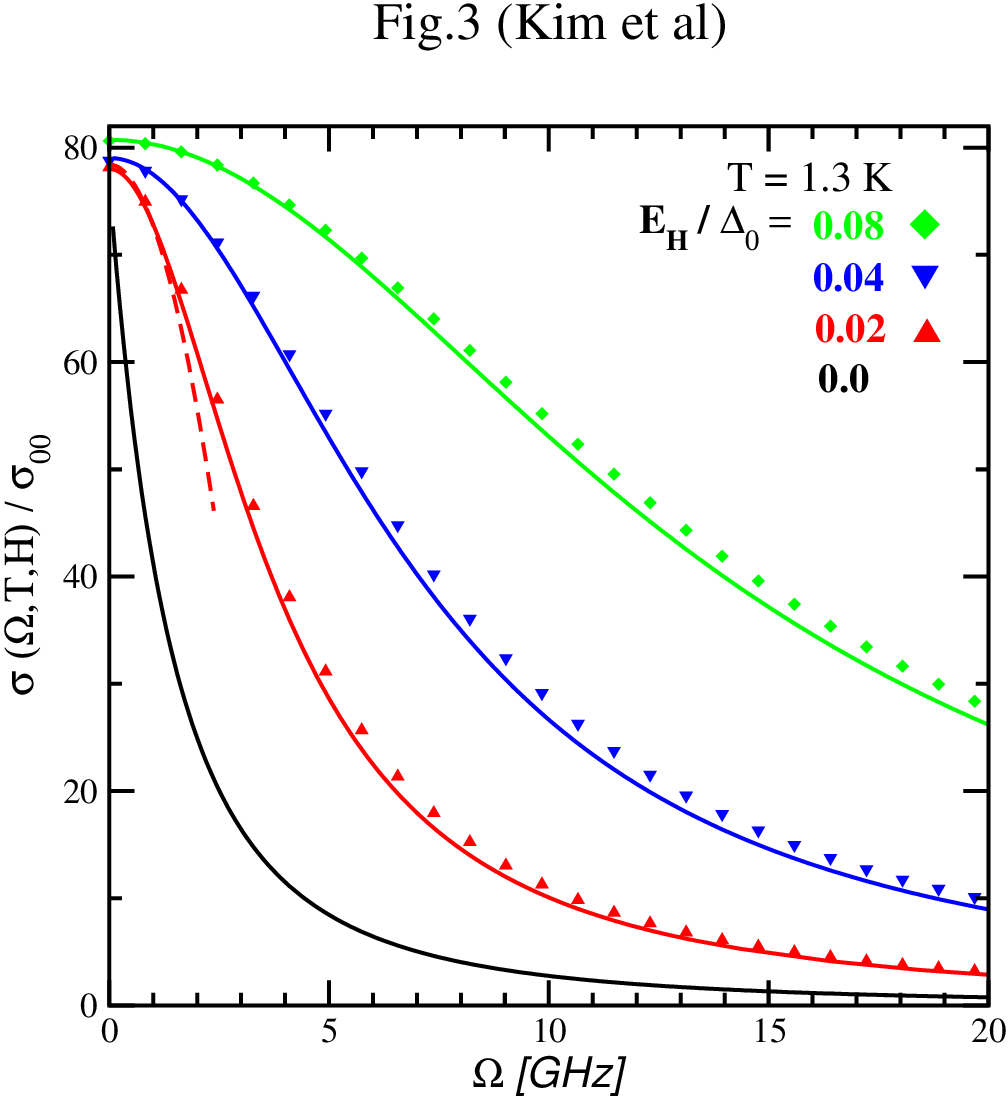}
\caption{ Microwave conductivity $\sigma(T,\Omega,H)$ vs. $\Omega$
up to $20$GHz for $T=1.3$K. The solid curves are results of
the full calculations using ${\cal A}_{\sigma}(\Omega,\omega,\epsilon,H)$
with Eq.~(\ref{self_imp_H}).
From the bottom to the top curve,
$E_{H}/\Delta_{0}=0$, $0.02$, $0.04$, and $0.08$.
The symbols are results of Eq.~(\ref{micro_vortex}), which is derived
under an assumption of the weak scattering limit.
The dashed curve based on the analytic expression Eq.~(\ref{drude_like})
shows that the cusp disappears.
}
\end{center}
\end{figure}

{\it In conclusion}, we have considered the low $T$
microwave conductivity of a high purity
$d$-wave superconductor with and without a magnetic field.
We find that the cusp-like behavior
in the microwave conductivity 
arises because the impurity scattering rate is negligibly small at zero frequency and
increases linearly at low frequency: $\gamma(\omega)\simeq\gamma_{00}+\alpha\omega$. 
The low frequency microwave conductivity does not depend on
$\gamma_{00}$ but the slope $\alpha$ plays a significant role.
The steepness of the cusp is inversely proportional to $T$. 
In the vortex state, the impurity scattering is drastically changed, and
its zero frequency value is no longer negligible.
Moreover, it increases quadratically with
frequency.
We predict that this characteristic change in the scattering rate
eliminates the cusp in the microwave conductivity.
Drude-like behavior appears instead.
We hope our prediction inspires an experiment of the microwave
conductivity in the vortex state.

This work is supported by the Natural Sciences and Engineering Research Council of 
Canada (NSERC), by ICORE (Alberta), and by the Canadian Institute 
for Advanced Research (CIAR).  

\bibliographystyle{prl}

\begin{thebibliography}{1}
\bibitem{other_dist} There are other two-node distributions available. See
Ref.\cite{vekhter}. However, they cannot be used for comparison with
experiments because
the normalized conductivity depends a lot on the energy cutoff.
\end{thebibliography}

\end{document}